\documentclass[
reprint,amsmath,amssymb,aps,pra,floatfix,longbibliography]{revtex4-1}
\usepackage{graphicx}	
\usepackage{dcolumn}	
\usepackage{bm}            
\usepackage{color}

\begin{document}
\title{
Optical pumping 
of
matrix-isolated 
barium 
monofluoride:
dependence on
the orientation of 
the 
BaF
molecular axis
}
\author{D. Heinrich}
\author{Z. Corriveau}
\author{J. Perez Garcia}
\author{N.T. McCall}
\author{H.-M. Yau}
\author{R.L. Lambo}
\author{T. Chauhan}
\author{G.K. Koyanagi}
\author{A. Marsman}
\author{M.C. George}
\author{C.H. Storry}
\author{M. Horbatsch}
\author{E.A. Hessels} 
\email{hessels@yorku.ca}
\affiliation{Department of Physics and Astronomy, York University, Toronto,
Ontario M3J 1P3, Canada}

\date{\today} 

\begin{abstract}

Optical pumping of 
barium 
monofluoride
(BaF)
within 
a cryogenic
neon matrix is 
demonstrated.
Interestingly,
with an applied   
magnetic field 
of 
2~G, 
optical pumping
is found to be considerably 
more efficient
for 
a laser beam with
right-circular
polarization
compared to 
left-circular
polarization.
Calculations show that 
the
higher efficiency
is 
due to 
a constructive versus 
destructive
interference
and
is dependent on the 
orientation of the 
BaF
molecule
relative to the
magnetic field direction.
The effect
leads to 
orientation-dependent
optical pumping within the 
matrix.
As optical pumping is the 
first step used in 
our planned 
electron 
electric-dipole
moment
(eEDM)
measurement,
we intend to 
exploit this property 
to obtain the 
selection of molecular
orientations 
that is
required for 
an 
eEDM
measurement.

\end{abstract}

\pacs{Valid PACS appear here}
\maketitle

\section{introduction}

For many decades, 
the electron 
electric-dipole 
moment 
(eEDM)
has been measured 
\cite{roussy2023new,
acme2018improved,
Cairncross2017,
baron2013order,
eckel2013search,
kara2012measurement,
hudson2011improved,
heidenreich2005limit,
hudson2002measurement,
regan2002new,
commins1994improved,
cho1991search,
abdullah1990new,
murthy1989new,
cho1989tenfold,
lamoreaux1987new,
vold1984search,
player1970experiment,
gould1970search,
weisskopf1968electric,
stein1967electric,
angel1967observation,
ensberg1967experimental,
sandars1964electric}
with increasing accuracy.
Earlier measurements used 
thermal beams of atoms
\cite{regan2002new,
commins1994improved,
abdullah1990new,
gould1970search,
player1970experiment,
weisskopf1968electric,
angel1967observation,
stein1967electric,
sandars1964electric},
atoms in cells
\cite{murthy1989new,
lamoreaux1987new,
vold1984search,
ensberg1967experimental},
or
bulk
properties of solids
\cite{heidenreich2005limit}.
More modern 
measurements 
use the 
large effective electric
field of heavy polar molecules
\cite{acme2018improved,
eckel2013search,
kara2012measurement,
hudson2011improved,
baron2013order,
cho1991search,
cho1989tenfold}
or
molecular ions
\cite{roussy2023new,
Cairncross2017}
to achieve 
a higher precision.
These 
eEDM 
measurements 
test for the type of 
time-reversal-symmetry 
violation that is required
to understand the asymmetry
between the abundances of matter and antimatter
in the universe.
Surprisingly,
despite the increasing accuracy,
the measured
eEDM 
\cite{roussy2023new}
remains consistent with 
zero. 
The increasingly small
limits on 
the
eEDM
strongly restrict the possible 
beyond-the-Standard-Model
extensions 
that are needed 
to explain the 
matter-antimatter
asymmetry. 

The 
EDM$^3$
collaboration has proposed 
\cite{vutha2018orientation,
vutha2018oriented}
the use
of 
matrix-isolated 
polar molecules
(in particular, 
barium monofluoride 
(BaF)
embedded in solid 
Ar 
\cite{koyanagi2023accurate,
lambo2023calculationAr}
or 
Ne
\cite{lambo2023calculationNe})
as a method for 
dramatically increasing the number of 
molecules that can participate
in an
eEDM experiment,
which could allow for 
much smaller statistical uncertainty
for an
eEDM 
determination.
Such a method may also allow for reduced
systematic uncertainties, 
since the molecules are cold, 
are stationary,
have their orientations fixed by the matrix
(without the need for an external 
field),
and 
are confined to a small volume.

The 
EDM$^3$
collaboration has grown
\cite{li2023optical,
corriveau2024matrix}
cryogenic
BaF-doped
Ne 
solids 
that have the desired
density of 
BaF
molecules
\cite{corriveau2024matrix}
($\sim$$10^{10}$~BaF/mm$^3$,
or a 
BaF:Ne 
ratio of 
$\sim$$1$:$10^9$).
The solids are 
grown using a
Ne 
gas flow and 
a beam of 
BaF
molecules from 
a 
helium-buffer-gas-cooled
laser-ablation 
source 
that are both incident on
a cryogenic sapphire
substrate.
Effectively, 
the solid freezes 
the 
BaF
molecules
produced in an hour 
of operation of the 
source,
allowing for the 
continued study of this
hour-long sample
of molecules.
The substrate is 
placed 
20~cm away from the  
buffer-gas 
source
to give the required space
needed for planned separation
\cite{marsman2023large,marsman2023deflection,yau2024specular}
of the 
BaF
molecules
from the other 
laser-ablation 
products.

Here we demonstrate 
optical pumping
of our 
matrix-isolated
BaF
molecules,
which is the first step
\cite{vutha2018orientation}
required for an 
eEDM
measurement.
Counter-intuitively,
at our magnetic field of 
2~G,
this optical pumping 
shows a significantly 
stronger response
when using
right-circular 
polarization 
(RCP)
for our laser
compared
to 
left-circular
polarization
(LCP).
Our calculations
(Sect.~\ref{sect:OPfixed})
show that the
higher efficiency
is dependent on the
orientation of the 
BaF
molecule,
with different
effects for 
oppositely oriented
molecules.

This orientation dependence
of optical pumping will 
allow for the 
selection of molecular
orientations. 
The selection of molecular
orientations is 
essential for an 
eEDM 
measurement, 
as the effective electric
field experienced by the 
unpaired electron
(and therefore the 
eEDM 
of this electron)
in the 
BaF 
molecule is aligned
with the molecular orientation.
The present scheme
for molecular orientation selection
is unique
among 
eEDM 
measurements in that it does
not require an applied 
electric field. 

\section{Optical pumping of molecules with fixed orientation \label{sect:OPfixed}}

\begin{figure}
\includegraphics
[width=1.0\linewidth]
{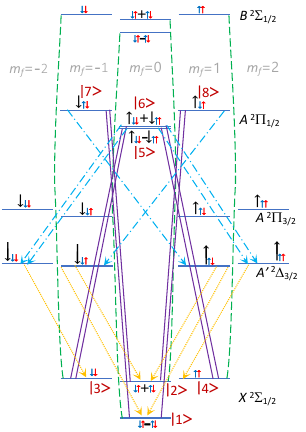}
\caption{
\label{fig:energyLevels}
(color online)
Relevant energy levels of the 
BaF
molecule (not to scale).
The arrows represent the 
component of the orbital
angular momentum 
(black),
electron spin
(blue),
and 
nuclear spin
(red)
along the 
internuclear
axis.
The solid 
(purple)
lines show the 
electric-dipole
decay (and excitation) 
paths for the 
$X\ ^2\Sigma_{1/2}$-$A\ ^2\Pi_{1/2}$
system.
Mixing of the 
$A\ ^2\Pi_{1/2}$
with the 
$B\ ^2\Sigma_{1/2}$
state allows for the additional 
electric-dipole 
decay and excitation routes 
shown as dashed 
(green) lines.
In a neon matrix, 
the 
$A\ ^2\Pi_{1/2}$
state is found to 
decay nonradiatively to the 
$A^\prime\ ^2\Delta_{3/2}$
state, 
as shown by the 
dot-dash
(light blue)
lines, 
with subsequent 
radiative decay 
(dotted,
orange lines).
Note that the 
more complicated decay through 
the 
dot-dash
and 
dotted line
combination
have the same starting 
and ending points as the 
direct radiative decay 
(solid lines).
}
\end{figure}

In this section, 
we consider 
optical pumping of
BaF
molecules 
with fixed orientations
due to 
excitation 
from 
the 
$X\,^2\Sigma_{1/2} (v$$=$$0)$
ground state to the 
$A\,^2\Pi_{1/2} (v$$=$$0)$
excited state
with 
circularly polarized laser light.
These states 
each have four hyperfine states,
as shown
in 
Fig.~\ref{fig:energyLevels}.
The purple solid lines in the 
figure are the allowed transitions
for this excitation 
(with the 
lines going upward and to the right
and left
being driven by 
right-
and 
left-circularly-polarized
light, 
respectively. 
The selection rules for the 
electric-dipole
transitions
for these
molecules of fixed orientation are 
determined by the projection of the 
orbital, 
electron-spin,
and
nuclear-spin
angular momenta 
($m_l$,
$m_s$,
and
$m_i$; 
which sum to 
$m_f$)
along the 
internuclear 
axis.
These projected angular momenta
are represented by the arrows 
in 
Fig.~\ref{fig:energyLevels}.

The figure reveals a problem for
optical pumping of
these 
non-rotating
molecules, 
namely that the 
$X\,^2\Sigma_{1/2} |m_f|$$=$$1$
states
together with the 
$A\,^2\Pi_{1/2} m_f$$=$$0$
states form a closed system
(with the 
$X\,^2\Sigma_{1/2} m_f$$=$$0$
and 
$A\,^2\Pi_{1/2} |m_f|$$=$$1$
states 
forming a second closed system).
Thus, 
it would appear that only half
of the molecules can be affected
by optical pumping.
In reality, 
this is not the case,
as the 
$A\,^2\Pi_{1/2}$
and
$B\,^2\Sigma_{1/2}$
states are mixed
\cite{bernard1990studiesII,
bernard1989bariumhydride}
by the 
fine-structure 
interaction,
leading to 
perturbation 
of the 
$A\,^2\Pi_{1/2}$
state:

\begin{eqnarray}
\label{eq:Amix}
&&a_\Sigma            \,|^2\Sigma_{1/2}\rangle
 +\sqrt{1-a_\Sigma^2} \,|^2\Pi_{1/2}   \rangle     
\end{eqnarray}
and a similar perturbation for the 
$B\,^2\Sigma_{1/2}$
state:
\begin{eqnarray}
\label{eq:Bmix}
&&-a_\Sigma            \,|^2\Pi_{1/2}   \rangle
 + \sqrt{1-a_\Sigma^2} \,|^2\Sigma_{1/2}\rangle,     
\end{eqnarray}
with 
$a_\Sigma=-0.20$.
The admixture allows for additional 
electric-dipole-allowed 
excitation
and decay paths 
(green dashed lines in 
Fig.~\ref{fig:energyLevels})
and allows
the population from all
four ground 
hyperfine
states to be optically pumped into a single state
($|3\rangle$
or 
$|4\rangle$
of 
Fig.~\ref{fig:energyLevels}
for 
LCP
or
RCP,
respectively).

The 
$m_f$ 
levels shown in 
Fig.~\ref{fig:energyLevels} 
use the 
internuclear 
axis as the quantization 
axis. 
Molecules that are
oriented parallel
(or 
antiparallel)
to the propagation direction
$\vec{k}$
for a laser beam
can be effectively optically 
pumped using 
right-
or
left-circularly-polarized
light
with  
$X\,^2\Sigma_{1/2} m_f$$=$$-1$
or 
$+1$
becoming a 
dark state.
However, 
molecules at other orientations 
do not have a dark state 
for
these circularly polarized
laser beams.
As a result,
the efficiency of optical 
pumping drops off quickly
for molecular orientations 
away from the 
direction 
$\vec{k}$
of the laser light.

A further complication occurs when 
a 
dc 
magnetic field is applied
(with its direction parallel 
or 
antiparallel 
to 
$\vec{k}$).
For molecules not oriented along 
$\vec{k}$,
the magnetic field mixes the 
hyperfine
states
of 
Fig.~\ref{fig:energyLevels},
which also has an effect on the 
presence of dark states.

\section{Density-matrix calculations}

To analyze the  
optical pumping,
we numerically
integrate the 
density matrix equations
for the 
eight-level
system 
(the 
four
hyperfine 
states of the 
$X\,^2\Sigma_{1/2} (v$$=$$0)$
ground state 
and the 
four 
hyperfine 
states of the 
$A\,^2\Pi_{1/2} (v$$=$$0)$
excited state
(including the admixture of 
Eq.~(\ref{eq:Amix})).
The 
Hamiltonian 
for the 
system has the form
\begin{eqnarray}
\label{eq:hamiltonian}
H=
\begin{pmatrix}
H_g   &H_{ge}\\
H_{ge}^*&H_e     \\
\end{pmatrix},
\end{eqnarray}
where
\begin{eqnarray}
\label{eq:Hg}
H_g=
\begin{pmatrix}
E_1                 &\mu_1 B_0        &\mu_2 B_-           &\mu_2 B_+           \\
\mu_1 B_0        &E_2                 &\mu_3 B_-           &-\mu_3 B_+            \\
-\mu_2 B_+        &-\mu_3 B_+        &E_3\!-\!\mu_4 B_0        &0                       \\
-\mu_2 B_-        &\mu_3 B_-       &0                       &E_3\!+\!\mu_4 B_0        \\
\end{pmatrix}, 
\end{eqnarray}
and
\begin{eqnarray}
\label{eq:He}
H_e=
\begin{pmatrix}
E_5                 &\mu_5 B_0        &\mu_6 B_-           &\mu_6 B_+        \\
\mu_5 B_0        &E_6                 &-\mu_6 B_-            &\mu_6 B_+        \\
-\mu_6 B_+        &\mu_6 B_+       &E_7\!+\!\mu_5 B_0        &0                    \\
-\mu_6 B_-        &-\mu_6 B_-        &0                       &E_7\!-\!\mu_5 B_0     \\
\end{pmatrix}.
\end{eqnarray}
In the
Hamiltonians
the 
first two 
rows and columns
represent 
odd and even
$m_f$$=$$0$
states 
$|0^\pm\rangle$
and 
the third
and fourth are 
the 
$m_f$$=$$-1$
and
$+1$
states
$|\!\!\pm\!\!1\rangle$,
where the 
fixed 
internuclear 
axis
is used as the 
quantization axis.
The energies
$E_i$
come from 
measurements and 
an effective 
Hamiltonian for 
the 
hyperfine 
interactions 
\cite{denis2022benchmarking},
while the 
magnetic-field 
terms 
come from 
measurements and 
effective
Zeeman 
Hamiltonians
\cite{Steimle2011MolecularBeam,
cahn2014zeeman}.
The quantities needed for these
Hamiltonians
(including the spherical tensor
components for the 
applied 
dc
magnetic field
$\vec{B}$)
are listed in 
Table~\ref{table:parameters}.

\begin{table}[b!]
\begin{ruledtabular}
\caption{\label{table:parameters} 
Parameters used to determine
the energy,
Zeeman 
shift
and 
dipole matrix elements of the 
BaF molecule.
}
\begin{tabular}{lll}
parameter&value&source
\\
\hline
$\mu_1$        &$\mu_B(g_\parallel+g_I)/2$$=$$1.0024\mu_B$      &\cite{cahn2014zeeman}\\
$\mu_2$        &$\mu_B(g_\perp+g_I)/2$$=$$0.9994\mu_B$          &\cite{cahn2014zeeman}\\
$\mu_3$        &$\mu_B(g_\perp-g_I)/2$$=$$0.9966\mu_B$          &\cite{cahn2014zeeman}\\
$\mu_4$        &$\mu_B(g_\parallel-g_I)/2      $$=$$0.9996\mu_B$&\cite{cahn2014zeeman}\\
$\mu_5$        &$\mu_B(g_S-2g_L^\prime)/2      $$=$$ 0.021\mu_B$ &\cite{Steimle2011MolecularBeam}\\
$\mu_6$        &$-g_l^\prime \mu_B/2            $$=$$ 0.190\mu_B$&\cite{Steimle2011MolecularBeam}\\
$E_1/h$        &$-A_{X}^{\parallel}/4-A_{X}^{\perp}/2$$=$$-49.68775$~MHz&\cite{denis2022benchmarking}\\
$E_2/h$        &$-A_{X}^{\parallel}/4+A_{X}^{\perp}/2$$=$$ 13.82125$~MHz&\cite{denis2022benchmarking}\\
$E_3/h$        &$A_{X}^{\parallel}/4$$=$$ 17.93325$~MHz          &\cite{denis2022benchmarking}\\
$E_5/h\!-\!f_0$&$-A_{A}^{\parallel}/4-A_{A}^{\perp}/2$$=$$-16.45$~MHz                                  &\cite{denis2022benchmarking}\\
$E_6/h\!-\!f_0$&$-A_{A}^{\parallel}/4+A_{A}^{\perp}/2$$=$$-12.87$~MHz                                  &\cite{denis2022benchmarking}\\
$E_7/h\!-\!f_0$&$A_{A}^{\parallel}/4$$=$$ 14.66$~MHz                                  &\cite{denis2022benchmarking}\\
$f_0$          &348.66~THz                                &\cite{Steimle2011MolecularBeam}\\
$\tau_a$,                            
$\tau_b$       &57.0(3), 41.7(3)~ns                              &\cite{aggarwal2019lifetime,
                                                                        berg1998lifetime,
                                                                        berg1993lifetime}\\ 
$a_\Sigma$,  
$a_\Pi$        &$-0.20$,$\sqrt{1-a_\Sigma^2}$$=$0.98             &\cite{bernard1990studiesII,
                                                                        bernard1989bariumhydride}\\
$\omega_{
\rm ag}$,
$\omega_{
\rm bg}$       &$2\pi f_0$,$2\pi$(420.91~THz)                    &\cite{bernard1992laser,
                                                                        rockenhauser2023absorption}\\
$d_{\rm ag}$   &$[3\pi\epsilon_0 \hbar c^3/
                 \tau_{\rm a}/\omega_{\rm ag}^3
                 ]^{1/2}$$=$$2.35~e\,a_0$                       &\\
$d_{\rm bg}$   &$[3\pi\epsilon_0 \hbar c^3/
                 \tau_b/\omega_{\rm bg}^3
                 ]^{1/2}$$=$$2.07~e\,a_0$                       &\\  
$B_0,B_\pm$    &$B_0$$=$$B_z$,
                $B_\pm$$=$$\mp(B_x \pm i B_y)/\sqrt{2}$          &\\
$E_0,E_\pm$    &$E_0$$=$$E_z$,
                $E_\pm$$=$$\mp(E_x \pm i E_y)/\sqrt{2}$          &\\             
$d_\Pi$,
$d_\Sigma$     &$a_\Sigma^2 d_\Pi^2+a_\Pi^2 d_\Sigma^2$$=$
                $d_{\rm bg}^2,a_\Pi^2 d_\Pi^2+a_\Sigma^2
                 d_\Sigma^2$$=$$d_{ag}^2$                        &\\
$d_\parallel$  &$-a_\Sigma d_\Sigma   $$=$0.41~$e\,a_0$         &\\
$d_\perp$      &$a_\Pi d_\Pi/\sqrt{2} $$=$1.63~$e\,a_0$         &\\
\end{tabular}
\end{ruledtabular}
\end{table}

The interaction 
Hamiltonian
$H_{ge}$
results from the 
electric-dipole 
interaction
with the 
oscillating
electric field
of the 
optical-pumping 
laser:
\begin{eqnarray}
\label{eq:Heg}
&&H_{ge}=
\begin{pmatrix}
-d_\parallel E_0       &0       &-d_\perp E_-              &-d_\perp E_+       \\
0       &-d_\parallel E_0       &-d_\perp E_-              &d_\perp E_+       \\
d_\perp E_+           &d_\perp E_+           &-d_\parallel E_0           &0                  \\
d_\perp E_-           &-d_\perp E_-           &0                           &-d_\parallel E_0          \\
\end{pmatrix}.
\end{eqnarray}
The electric dipole moments 
($d_\perp$
and
$d_\parallel$)
in this 
Hamiltonian
can be deduced from the mixing 
(Eq.~(\ref{eq:Amix}))
and the measured lifetimes of the 
mixed states,
as detailed in 
Table~\ref{table:parameters}.

The full set of 
density-matrix
equations includes the 
Hamiltonian 
of 
Eq.~(\ref{eq:hamiltonian})
and 
radiative 
decay 
of the 
$A\,^2\Pi_{1/2} (v$$=$$0)$
state.
These 
density-matrix
equations have the same form
as those in 
Eq.~(1)
of 
Ref.~\cite{marsman2023deflection},
including
\cite{cardimona1983spontaneous,
marsman2012shifts}
quantum-mechanical
interference terms
for spontaneous
emission. 
The lifetime for this 
decay has been measured
\cite{aggarwal2019lifetime,
berg1998lifetime},
with the weighted average giving 
57.0(3)~ns.
The branching ratios from
each of the four 
$A\,^2\Pi_{1/2}(v=0)$
states
to each of the four 
$X\,^2\Sigma_{1/2}$
states are proportional
to the squares of the 
matrix elements in 
Eq.~(\ref{eq:Heg}).
Numerical integration of the 
density-matrix 
equations follows methods
similar to
those used in 
Refs.~\cite{marsman2023large}
and
\cite{marsman2023deflection}.

We ignore the smaller 
branching ratios 
(of approximately
3.5\%
\cite{hao2019high})
for radiative
decay to
higher-$v$
levels. 
In a matrix, 
decay to a 
higher-$v$
state will 
follow the same 
branching ratios 
(with respect to 
hyperfine states)
as the decay to 
$v$$=$0
and will 
be followed by
a 
matrix-assisted
vibrational 
relaxation to 
the 
$v$$=$0
state.
This relaxation is 
also not expected to 
change the hyperfine
state, 
as the matrix interactions
are primarily electric
and thus do not 
affect electron
or nuclear
spin.

\begin{figure}
\includegraphics
[width=0.7\linewidth]
{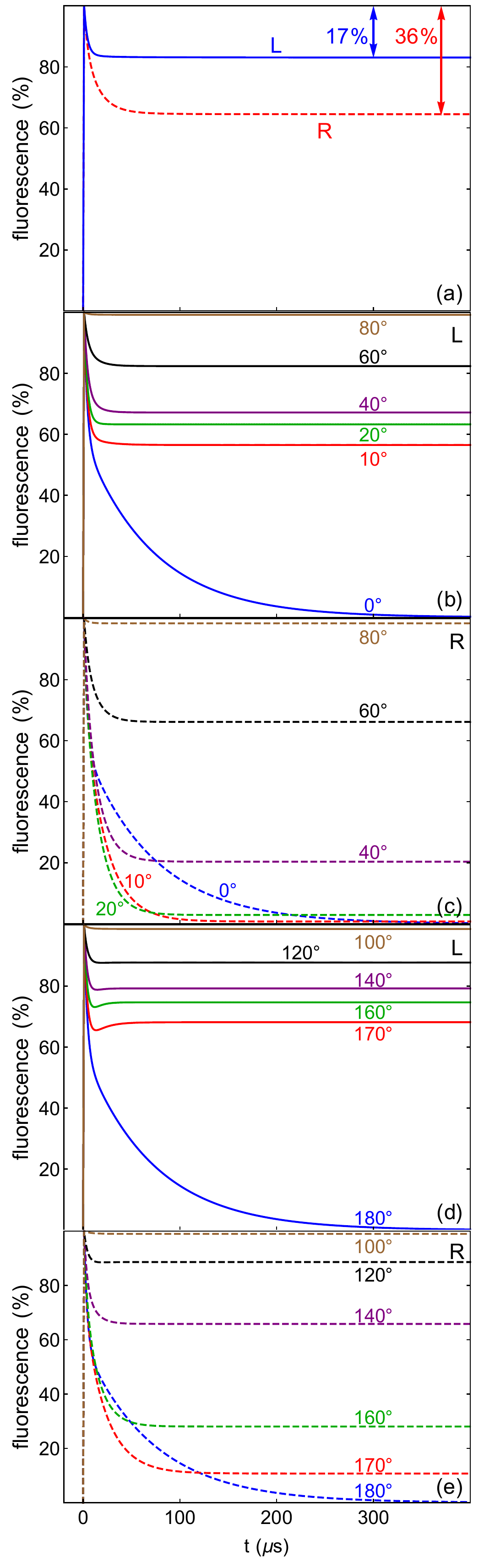}
\caption{
\label{fig:fluorVsTime}
(color online)
The fluorescence versus
time for an ensemble of 
BaF
molecules
being optically pumped
using 
circularly-polarized
light 
(with 
$|\vec{E}|$
of
1.77~V/cm)
in a magnetic
field of 
$B$$=$2~G.
Panel 
(a),
which 
shows the total 
fluorescence for 
RCP 
and 
LCP,
reveals that optical 
pumping is more efficient
for 
RCP.
The fluorescence 
for various angles
$\theta$
of the molecular orientation
relative to 
$\vec{B}$
is shown in 
panels
(b)
through
(e),
and these show 
the increased efficiency
of 
RCP
at small angles.
}
\end{figure}

Our simulations begin with
an ensemble of molecules
whose orientations,
$\vec{o}$,
are 
isotropically distributed,
with
equal populations in states
$|1\rangle$
through 
$|4\rangle$ 
of 
Fig.~\ref{fig:energyLevels}.
The total fluorescence 
from this ensemble 
is calculated, 
and the reduction in 
fluorescence versus 
time 
(shown in 
Fig.~\ref{fig:fluorVsTime}(a))
is a direct indication
of optical pumping.
We define the efficiency
of optical pumping as the 
fractional reduction in
fluorescence,
as our modeling shows that 
this reduction is due to 
a buildup of 
dark-state
population.
From the figure, 
one can see that 
optical pumping is
predicted to be more efficient
(at an applied magnetic field
of 
$\vec{B}$
of 
2~G 
that is parallel
to 
the laser propagation direction
$\vec{k}$)
for 
right circular polarization
(RCP)
than for 
left circular polarization
(LCP). 

\begin{figure}
\includegraphics
[width=1.0\linewidth]
{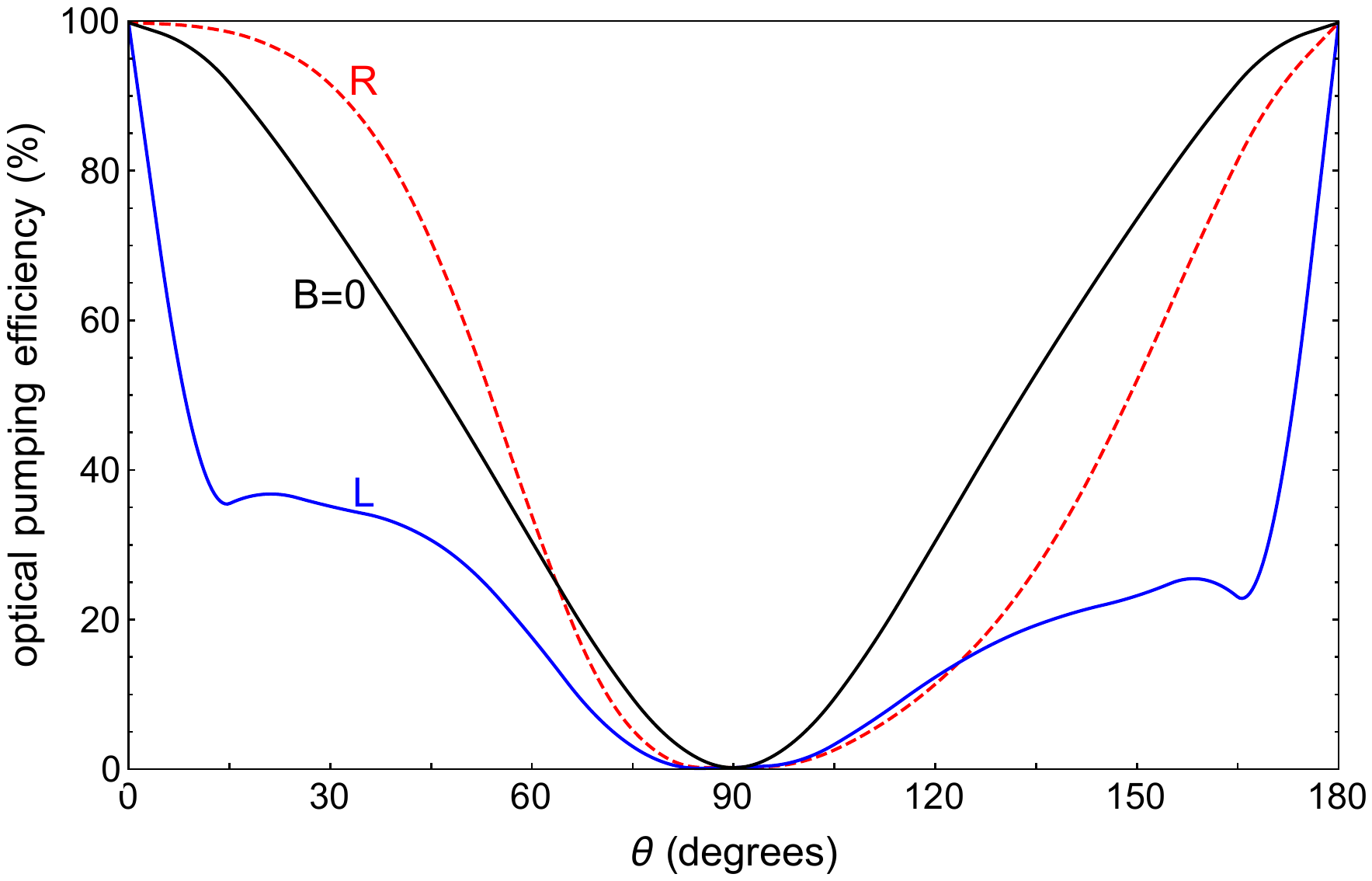}
\caption{
\label{fig:OptPumpEfficiency}
(color online)
The 
calculated
efficiency
of optical pumping
for 
$B$$=$2~G
(for 
RCP 
and
LCP)
versus 
the angle between 
the 
BaF 
molecular
orientation
and 
$\vec{B}$.
Note the 
increased optical
pumping for 
small angles
and the decrease 
for angles 
approaching
180$^\circ$.
The 
symmetric result 
for
$B$$=$0
is shown for comparison.
}
\end{figure}

\begin{figure}
\includegraphics
[width=0.85\linewidth]
{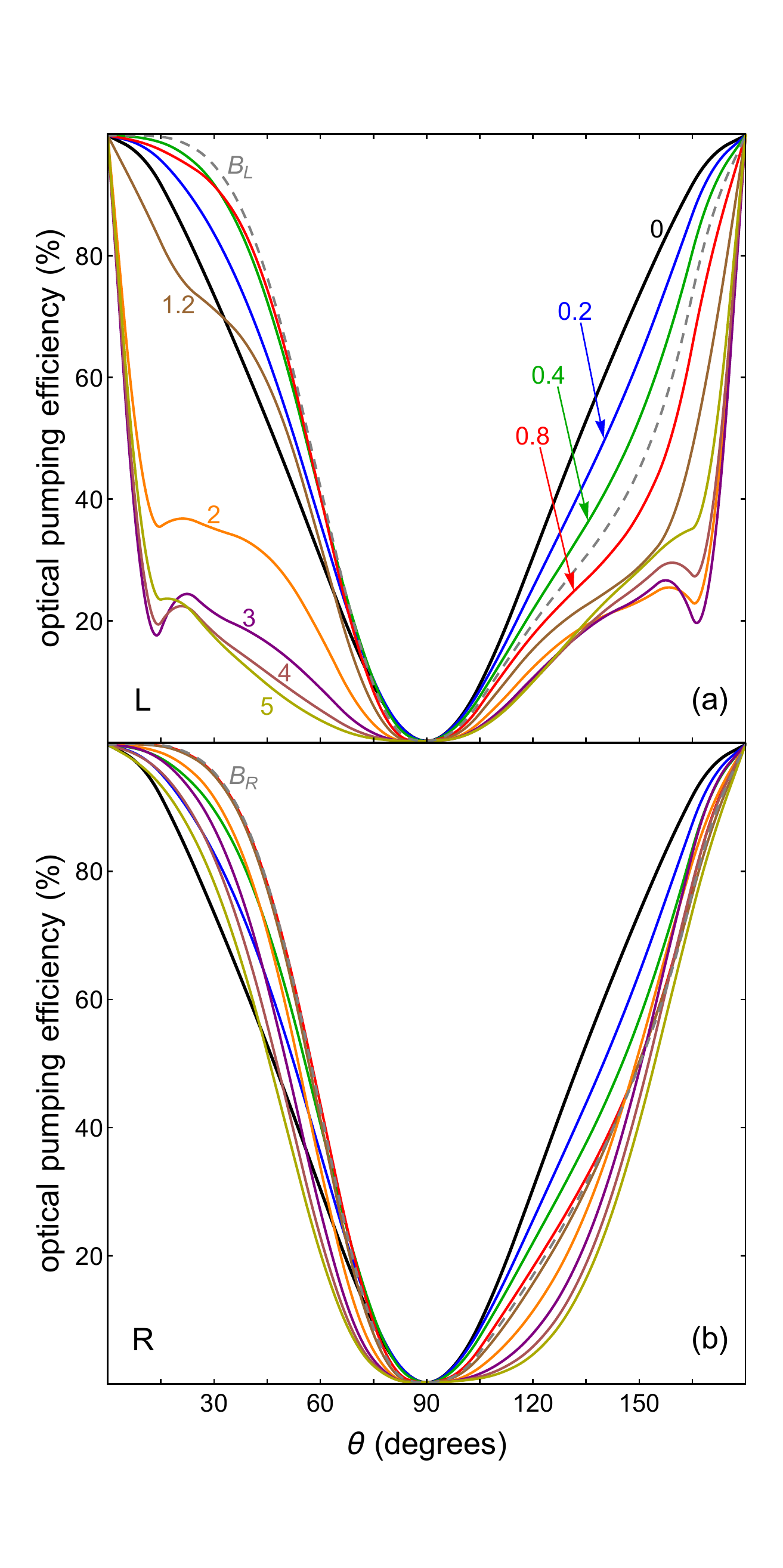}
\caption{
\label{fig:OptPumpDiffB}
(color online)
The efficiency
of optical pumping
for 
a range of 
magnetic fields
(0,
0.2,
0.4,
0.8,
1.2,
2,
3,
4,
and
5~G,
as labeled 
for 
LCP
in
(a)
and
color-coded
for
the same fields
for
RCP
in
(b)).
The dashed lines show 
the magnetic fields
$B_{\rm L}$
and
$B_{\rm R}$
of 
Eqs.~(\ref{eq:BL})
and
(\ref{eq:BR}),
for which 
optical pumping is efficient
over a larger range of 
angles near
$\theta$$=$$0$
}
\end{figure}

To explore the origin of this
surprising effect, 
we calculate the 
optical pumping efficiency
for each circular polarization
for specific molecular 
orientations, 
$\vec{o}$.
We find that the symmetry 
of our Hamiltonian leads to
identical efficiencies as 
a function of 
$\phi$ 
(the azimuthal angle
of
$\vec{o}$
with respect to
$\vec{B}$).
The dependence on 
$\theta$
(the angle 
between 
$\vec{o}$
and
$\vec{B}$)
can be seen in 
Fig.~\ref{fig:fluorVsTime}(b)
through 
(e),
and is summarized
in 
Fig.~\ref{fig:OptPumpEfficiency}.
Note that optical pumping 
is 
100\% 
efficient 
for
$\theta$
equals
$0^{\circ}$
or
$180^{\circ}$
for both 
LCP 
and 
RCP.
For 
LCP,
the optical pumping 
efficiency drops off
quickly as 
$\theta$
increases from
$0^{\circ}$
and as it decreases from 
$180^{\circ}$,
but for 
RCP,
it 
drops off much
more slowly 
as 
$\theta$
increases from
$0^{\circ}$.
For example,
Fig.~\ref{fig:OptPumpEfficiency}
shows that optical pumping has a 
more than 
50\%
higher efficiency for 
RCP
(compared to 
LCP)
for 
$10^\circ<\theta<35^\circ$.
The higher efficiency
for a range of orientations
near
$\theta$$=$$0^{\circ}$
(and,
to a lesser 
extent,
for angles 
near
$\theta$$=$$180^{\circ}$)
is the reason that the 
ensemble 
(Fig.~\ref{fig:fluorVsTime}(a))
results in a higher overall
efficiency for 
RCP.
Fig.~\ref{fig:OptPumpDiffB}(b) 
shows that the
increased efficiency is significant
for a range of magnetic field strengths
near
2~G.

These 
density-matrix 
calculations show that
for 
RCP
there are more 
optically pumped
molecules
that are nearly parallel with 
$\vec{B}$ 
than those nearly antiparallel.
This effect will be key to 
performing an
eEDM
experiment, 
as it will allow for the 
preparation of a sample 
of 
optically pumped
molecules that are predominately
oriented near to the direction
of the magnetic field.
The use of 
LCP
would prepare a sample that
does not have this predominant
orientation, 
which could then be used as
a control.
Further controls would 
result from reversal of 
$\vec{B}$
and of
$\vec{k}$.
A more complete description
of a planned 
eEDM 
measurement using 
this method of preferred
optical pumping will be the 
subject of a future publication.

\section{A simplified model for understanding the effect}

High efficiency for optical 
pumping results when 
dark 
(or 
nearly-dark)
states
are present. 
In this section, 
we use perturbation theory
and expansions in small parameters
to explain the presence of dark
states 
for 
molecular orientations
$\theta$
near 
$0^{\circ}$ 
or
$180^{\circ}$.
These results are less precise
than those of the previous section,
but allow for a physical understanding
of the effects predicted by the 
full
density-matrix 
treatment.

Looking at
$H_{\rm g}$
of 
Eq.~(\ref{eq:Hg}),
the large 
($\sim$65~MHz)
separation
between 
$E_1$
and the other diagonal
elements  
compared to the 
small 
($<$3~MHz
for 
$B$=2~G)
values
for the 
off-diagonal
elements
makes this state
nearly unmixed.
To an accuracy of 
a few percent 
considered in this 
section, 
only the three other
states mix.
The applied
dc
magnetic field
$\vec{B}$$=$$(0$,$0$,$B)$
(where
$\vec{k}$
is assumed to be 
in the direction of 
the 
$z$
axis)
transforms into 
$(-B\sin\theta$,
$0$,
$B\cos\theta)$
in the molecular frame.
In terms of a 
spherical tensor used
in 
Eqs.~(\ref{eq:Hg}),
this becomes
$(B_-,B_0,B_+)$
$=$
$(-B\sin\theta/\sqrt{2}$,
$B\cos\theta$,
$B\sin\theta/\sqrt{2})$.
Also ignoring the small
nuclear magnetic moment,
the 
submatrix 
of 
Eq.~(\ref{eq:Hg})
for the three mixed 
states is
\begin{eqnarray}
\begin{pmatrix}
E_2&          
-V&
-V\\
-V&
E_2+\Delta_-&
0\\
-V&
0&
E_2+\Delta_+
\end{pmatrix},
\end{eqnarray}
where
$V$$=$$\frac{1}{\sqrt{2}}\mu_B B\sin\theta$
and
$\Delta_\pm$$=$$E_3-E_2\pm \mu_B B\cos\theta$.
For small 
$\theta$$\lesssim$20$^{\circ}$,
the perturbation 
$V$
is weak
($\lesssim$0.1~MHz)
compared to the 
energy differences
$\Delta_\pm$,
which are 
$\gtrsim$2~MHz.
Thus, 
first-order 
perturbation theory
is sufficient at the accuracy
considered in this section.
To first order,
the 
eigenstates 
are
\begin{eqnarray}
|1\rangle&=&|0^-\rangle_{\rm g},
\nonumber
\\
|2\rangle&=&|0^+\rangle_{\rm g}+\epsilon_-|\!-\!1\rangle_{\rm g}+\epsilon_+|\!+\!1\rangle_{\rm g},
\nonumber
\\
|3\rangle&=&|\!-\!1\rangle_{\rm g}-\epsilon_-|0^+\rangle_{\rm g},
\nonumber
\\
|4\rangle&=&|\!+\!1\rangle_{\rm g}-\epsilon_+|0^+\rangle_{\rm g},
\end{eqnarray}
where
$\epsilon_\pm$$=$$V/\Delta_\pm$.
For the excited states, 
the 
Hamiltonian
(Eq.~(\ref{eq:He}))
has the same structure,
but the smaller 
values of
$\mu$
make the unperturbed states
($|5\rangle$$=$$|0^-\rangle_{\rm e}$,
$|6\rangle$$=$$|0^+\rangle_{\rm e}$,
$|7\rangle$$=$$|\!-\!1\rangle_{\rm e}$,
and
$|8\rangle$$=$$|\!+\!1\rangle_{\rm e}$)
an accurate approximation
at the level of a few percent.

For the interaction
Hamiltonian
(Eq.~(\ref{eq:Heg})),
we take the electric
field of the laser beam
\begin{eqnarray}
\vec{E}=
{\rm Re}
[
|\vec{E}|
(\hat{x}\mp i\hat{y})
e^{i\omega t}
]    
\end{eqnarray}
(where the top sign
corresponds to 
RCP
and the bottom to 
LCP),
and transform it into
the molecular frame
and 
express it in terms of 
a spherical tensor:
\begin{eqnarray}
\begin{pmatrix}
E_-\\
E_0\\
E_+
\end{pmatrix}
&=&
|\vec{E}|
\begin{pmatrix}
\frac{1}{\sqrt{2}} (\cos\theta\mp 1) \\
\sin\theta\\
-\frac{1}{\sqrt{2}} (\cos\theta\pm 1) 
\end{pmatrix}
e^{\mp i\phi}
e^{i\omega t}.
\end{eqnarray}
Thus,
$H_{\rm ge} / |\vec{E}|$
(to within a phase)
is
\begin{eqnarray}
\begin{pmatrix}
d_\parallel \sin\theta
&
0
&
\frac{d_\perp(\cos\theta\mp 1)}{\sqrt{2}}
&
\frac{-d_\perp(\cos\theta\pm 1)}{\sqrt{2}}
\\
0
&
d_\parallel \sin\theta
&
\frac{d_\perp(\cos\theta\mp 1)}{\sqrt{2}}
&
\frac{d_\perp(\cos\theta\pm 1)}{\sqrt{2}}
&
\\
\frac{d_\perp(\cos\theta\pm 1)}{\sqrt{2}}
&
\frac{d_\perp(\cos\theta\pm 1)}{\sqrt{2}}
&
d_\parallel \sin\theta
&
0
\\
\frac{-d_\perp(\cos\theta\mp 1)}{\sqrt{2}}
&
\frac{d_\perp(\cos\theta\mp 1)}{\sqrt{2}}
&
0
&
d_\parallel \sin\theta
\end{pmatrix}.
\nonumber
\\
\end{eqnarray}
Using the perturbed
states
$|1\rangle$
to
$|8\rangle$
as a basis set 
leads to
\begin{eqnarray}
\begin{pmatrix}
\!
d_\parallel \sin\theta
&
0
&
\!\!\!\!\!\!
\frac{d_{\!\perp}\!(\cos\!\theta\mp 1)}{\sqrt{2}}
&
\!\!\!\!\!\!
\frac{-d_{\!\perp}\!(\cos\!\theta\pm 1)}{\sqrt{2}}
\\
\!
\begin{matrix}
\frac{ ( \epsilon_{-}\! - \epsilon_{+}\! ) d_{\perp}\! \!\cos\theta } {\sqrt{2}}
\\
{\scriptstyle \pm} 
\frac{ (\epsilon_{+}\! + \epsilon_{-\!})  d_{\perp} } {\sqrt{2}}
\end{matrix}
&
\!\!\!\!\!\!
\begin{matrix}
d_\parallel \sin\theta
{\scriptstyle +}
\\
\frac{ ( \epsilon_{-}\!+ \epsilon_{+}\! ) d_{\perp}\!\! \cos\theta }  {\sqrt{2}}
\\
{\scriptstyle \pm}
\frac{(\epsilon_{-}\! - \epsilon_{+}\! )d_{\perp} } {\sqrt{2}}
\end{matrix}
&
\!\!\!\!\!\!
\frac{d_{\!\perp}\!(\cos\!\theta\mp 1)}{\sqrt{2}}
&
\!\!\!\!\!\!
\frac{d_{\!\perp}\!(\cos\theta\pm 1)}{\sqrt{2}}
\\
\!
\frac{d_{\!\perp}\!(\cos\!\theta\pm 1)}{\sqrt{2}}
&
\!\!\!\!\!\!
\frac{d_{\!\perp}\!(\cos\!\theta\pm 1)}{\sqrt{2}}
&
\!\!\!\!\!\!
\begin{matrix}
d_\parallel \sin\theta
{\scriptstyle -}
\\  
\frac{\epsilon_-d_{\!\perp}\!(\cos\!\theta\mp 1)}{\sqrt{2}}
\end{matrix}
&
\!\!\!
\frac{\epsilon_- d_{\!\perp}\!(\cos\!\theta\pm 1)}{-\sqrt{2}}
\\
\!
\frac{d_{\!\perp}\!(\cos\!\theta\mp 1)}{-\sqrt{2}}&
\!\!\!\!\!\!
\frac{d_{\!\perp}\!(\cos\!\theta\mp 1)}{\sqrt{2}}
&
\!\!\!\!\!\!
\frac{\epsilon_+ d_{\!\perp}\!(\cos\theta\mp 1)}{-\sqrt{2}}
&
\!\!\!
\begin{matrix}
d_\parallel \sin\theta
{\scriptstyle -}   
\\ 
\frac{\epsilon_+ d_{\!\perp}\!(\cos\theta\pm 1)}{\sqrt{2}}
\end{matrix}
\end{pmatrix}.
\nonumber
\\
\  
\label{eq:HegApprox1to8}
\end{eqnarray}
Here, 
terms proportional to 
$\epsilon_\pm \sin\theta$
have been discarded as 
they scale as 
$\theta^2$
and contribute
at most at the 
few-percent 
level for 
$\theta$
near 
0$^\circ$
or
180$^\circ$.

Note that for 
RCP
(the top sign
in 
Eq.~(\ref{eq:HegApprox1to8}))
and small 
$\theta$,
row 4
(corresponding to state
$|4\rangle$)
has zeros
(to order 
$\theta^2$)
for the first
three elements.
The last element  
also gives zero
if
\begin{eqnarray}
d_\parallel \sin\theta
-
\frac{1+\cos\theta}{2}
\frac{d_{\perp} \mu_B B \sin\theta}{E_3-E_2+\mu_B B \cos\theta}
=
0.
\label{eq:cancelR}
\end{eqnarray}
For 
small
$\theta$
($\cos\theta$$\approx$$1$),
this equality holds 
(leading to destructive
interference for the two terms
in 
Eq.~(\ref{eq:cancelR}))
for a 
magic magnetic field of 
\begin{equation}
B_{\rm R}= 
\frac{d_\parallel}{d_\perp-d_\parallel}
\frac{E_3-E_2}{\mu_B}
\approx0.98~{\rm G}.   
\label{eq:BR}
\end{equation}
That is, 
at this field, 
$|4\rangle$
is a dark state
not only at 
$\theta=0$,
but also for 
a range of 
small
$\theta$.
For 
$\theta=180^{\circ}$,
optical pumping is still
efficient,
but,
for nearby angles, 
since
$\cos\theta \approx -1$,
the 
second term in 
Eq.~(\ref{eq:cancelR})
cannot cancel the first.

Similarly, 
for 
LCP
(the bottom sign in 
Eq.~(\ref{eq:HegApprox1to8})),
row~3
has zero elements for 
columns~1,
2,
and 4.
The element in 
column~3 
is zero if
\begin{eqnarray}
d_\parallel \sin\theta
-
\frac{1+\cos\theta}{2}
\frac{d_{\perp} \mu_B B \sin\theta}{E_3-E_2-\mu_B B \cos\theta}
=
0.
\label{eq:cancelL}
\end{eqnarray}
This equality holds
(for small
$\theta$)
for a 
magic magnetic field of 
 \begin{equation}
B_{\rm L}= 
\frac{d_\parallel}{d_\perp+d_\parallel}
\frac{E_3-E_2}{\mu_B}
\approx0.59~{\rm G}.
\label{eq:BL}
\end{equation}

Because both the denominator
and numerator in the 
second term of 
Eq.~(\ref{eq:cancelR}) 
increase with increasing
$B$, 
there is a nearly dark state
and enhanced optical pumping
for 
RCP
at small
$\theta$ 
for a large range of 
magnetic fields,
as can be seen in
Fig.~\ref{fig:OptPumpDiffB}(b).
The 
second term of
Eq.~(\ref{eq:cancelL})
increases quickly with
increasing
$B$,
and therefore there is a more
limited range for enhanced
small-$\theta$
optical pumping
for 
LCP,
as seen in 
Fig.~\ref{fig:OptPumpDiffB}(a).

\section{Observed optical pumping 
versus handedness of circular polarization}

\begin{figure}
\includegraphics
[width=0.85\linewidth]
{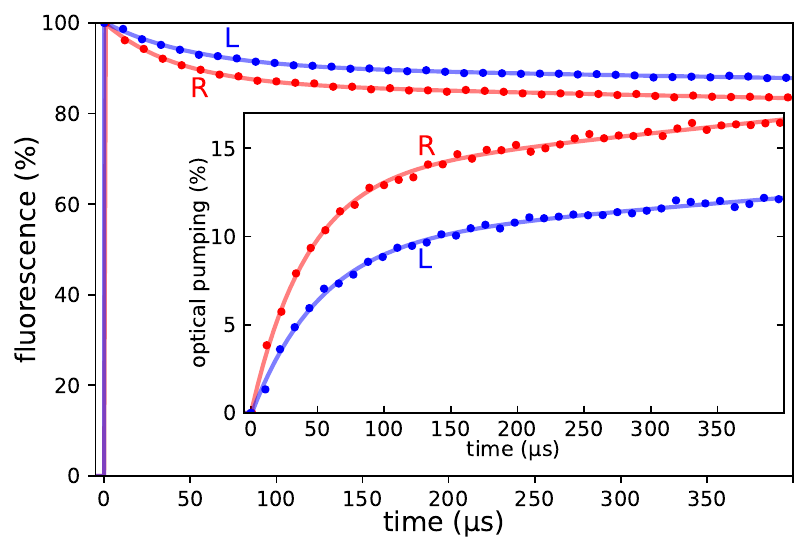}
\caption{
\label{fig:ExptOptPump}
(color online)
Experimental observation 
of optical pumping for
BaF
molecules
in a 
neon
matrix
with a magnetic
field of 
2~G.
The larger reduction in  
fluorescence  
for  
RCP
compared to 
LCP
indicates that 
RCP
is more efficient at 
optical pumping,
as shown in the inset.
}
\end{figure}

Figure~\ref{fig:ExptOptPump}
shows the observed fluorescence 
for 
optically pumping 
neon-matrix-isolated
BaF 
molecules
for 
$B$$=$2~G
using 
lasers with 
RCP 
and 
LCP.
An 
859.27-nm 
laser is used 
to optically pump
by driving the 
$X\ ^2\Sigma_{1/2}(v$$=$$0)$-to-$A\ ^2\Pi_{1/2}(v$$=$$0)$
transition.
As predicted
by our 
density-matrix
calculations
(Fig.~\ref{fig:fluorVsTime}(a)),
optical pumping 
with 
RCP
is found to 
be considerably more efficient
than with 
LCP.

Although the preference
for 
RCP 
over 
LCP
for optical pumping is 
clear in both 
Figs.~\ref{fig:fluorVsTime}(a)
and 
\ref{fig:ExptOptPump},
the details of the graphs differ.
The reasons for the differences
are almost certainly due to 
effects of the matrix that have
not been included in our 
density-matrix treatment.
The most obvious of these effects
is that the 
$A\ ^2\Pi_{1/2}$
state 
is found to 
decay 
nonradiatively
to the 
$A^\prime\ ^2\Delta_{3/2}$
state,
followed by spontaneous 
decay from this
state.
The spectrum of this decay 
was previously reported in 
Ref.~\cite{corriveau2024matrix},
where 
the same effect was observed
for the 
$B\ ^2\Sigma_{1/2}$
state.
For the 
$B\ ^2\Sigma_{1/2}$
state,
the 
fraction of decays that 
go through the
$A^\prime\ ^2\Delta_{3/2}$
state 
decreases dramatically as the
temperature of the solid is 
lowered by one kelvin.
A similar effect might be expected
for the 
$A\ ^2\Pi_{1/2}$
state
(allowing for direct decay 
with future upgrades of 
our apparatus),
but at our current base temperature
of approximately 
6~kelvin,
the vast majority decay 
through the 
$A^\prime\ ^2\Delta_{3/2}$
state.

Decay through the 
$A^\prime\ ^2\Delta_{3/2}$
state 
might be expected to completely scramble the 
optical-pumping process.
However, 
the 
nonradiative 
decay 
to  
$A^\prime\ ^2\Delta_{3/2}$
should be insensitive to 
spins, 
leading to the dominant
decay paths 
shown as
blue
dot-dash
lines in 
Fig.~\ref{fig:energyLevels}.
The 
electric-dipole-allowed
radiative decay paths from  
$A^\prime\ ^2\Delta_{3/2}$
to the ground state are shown
by the 
orange
dotted
lines.
Note that the 
combinations of the 
nonradiative
decay
(blue)
and 
radiative 
decay 
(orange)
follow the same 
paths as the 
direct radiative
decay from the 
$A\ ^2\Pi_{1/2}$
state
(purple solid lines
of 
Fig.~\ref{fig:energyLevels})
that were used for our 
density-matrix 
calculations.
As a result, 
the dominant decay paths 
for both processes are
identical.
The very fast 
nonradiative 
decay
does,
however, 
require a much higher
intensity laser,
and the data of 
Fig.~\ref{fig:ExptOptPump}
were taken at a 
2000-times 
higher 
laser-field 
amplitude than
the calculations of 
Fig.~\ref{fig:fluorVsTime}.

A small mixing coefficient
(0.14)
of the
$A^\prime\ ^2\Delta_{3/2}$
state
with the 
$A\ ^2\Pi_{3/2}$
state
\cite{bernard1990studiesII,
bernard1989bariumhydride}
(analogous to the 
mixing of 
Eq.~(\ref{eq:Amix}))
leads to other decay 
paths
that do not preserve
$m_s$,
which has a small 
scrambling effect 
on the
optical-pumping 
process.
Both this mixing 
coefficient and the 
one from 
Eq.~(\ref{eq:Amix})
may be affected by the 
matrix.
The
$A^\prime\ ^2\Delta_{3/2}$-$A\ ^2\Pi_{3/2}$
and
$A\ ^2\Pi_{1/2}$-$B\ ^2\Sigma_{1/2}$
spin-orbit 
mixing
are between states with the same 
value of 
$\Omega$,
as is required by the 
azimuthal 
symmetry about the 
internuclear axis.
In the matrix,
however, 
this 
azimuthal 
symmetry is
broken, and additional 
mixing is also possible.
All of these mixings
would be expected to reduce the 
efficiency of optical pumping.
Additionally, 
the 
nonradiative 
matrix-assisted
processes 
of decay to the 
$A^\prime\ ^2\Delta_{3/2}$
state
and from 
vibrationally
excited ground states is 
not well understood.
Finally, 
impurities in our solid
may cause stray magnetic fields
that would also affect the 
optical-pumping
process.

As a result of these 
poorly-understood 
matrix effects,
a complete modeling of
the optical pumping is 
not yet possible.
For the current work, 
we show that the observed effect 
of using 
RCP 
versus 
LCP 
for optical pumping 
matches the general behavior
of our model.
Expected upgrades to our apparatus 
will allow for lower temperatures,
where matrix effects are expected
to be strongly suppressed, 
and 
higher-purity
solids,
with the expectation that the 
resulting experimental observations
might be more completely modeled.

\section{conclusions}

We have demonstrated optical pumping 
of 
BaF
molecules in a 
cryogenic 
neon 
matrix, 
and have shown that 
the efficiency of optical pumping 
in a magnetic field
depends on the handedness of 
the circular polarization used.
Our calculations indicate that 
this effect can be explained 
and that 
optical-pumping
efficiency depends on the orientation
of the 
BaF
molecule within the matrix.
We intend to use this 
orientation-dependent 
effect for selecting molecular
orientations for performing 
an 
electron electric dipole
moment measurement.

\vspace{7mm}
\section*{Acknowledgments}
This work is supported by the 
Alfred P. Sloan Foundation,
the 
Gordon and Betty Moore Foundation, 
the 
Templeton 
Foundation
in conjunction with the 
Northwestern Center for Fundamental Physics, 
the 
Canada Foundation for Innovation,
the
Ontario Research Fund,
the 
Natural Sciences and Engineering Research Council
of Canada,
and 
York University. 
Computations for this work were supported by 
Compute Ontario 
and 
the 
Digital Research Alliance of Canada. 
The authors are grateful for 
extensive discussions with 
Amar Vutha 
and
Jaideep Singh,
which helped to guide this work.

\bibliography{BaF}

\end{document}